\documentclass[aps,prmaterials,amsmath,amssymb,reprint,superscriptaddress]{revtex4-1}
\usepackage{charter,graphicx,verbatim,threeparttable,float,xcolor,booktabs,textcomp,hyperref,bm}
\usepackage[version=4]{mhchem}

\newcommand{\tempapp}{T} 
\newcommand{\MScomp}{$M^0_{\text{net}}$} 
\newcommand{\MSexp}{$M_{\text{sat}}$} 
\newcommand{\Mspont}{$M_0$} 
\newcommand{\mueff}{\mu_{\text{eff}}}
\newcommand{\MUeff}{$\mueff{}$}
\newcommand{\GenSpinel}{$A_xC_{1-x}B_2$O$_4$}
\newcommand{\CurveFitName}{\texttt{scipy.optimize.curve\char`_fit}}

\begin{document}
\author{Joya\,A.\,Cooley}
\email{jcooley@fullerton.edu}
\affiliation{Department of Chemistry and Biochemistry, California State University, Fullerton, California 92834, United States}

\author{Gregor\,Dairaghi}
\affiliation{Department of Physics, Carleton College, Northfield, Minnesota 55057, United States}
\affiliation{Current Affiliation: Applied Physics Graduate Program, Northwestern University, Evanston, Illinois 60208, United States}

\author{Guy\,C.\,Moore}
\affiliation{Department of Materials Science and Engineering, University of California Berkeley, Berkeley, California 94720, United States}
\affiliation{Materials Science Division, Lawrence Berkeley National Laboratory, Berkeley, California 94720, United States}

\author{Matthew\,K.\,Horton}
\affiliation{Department of Materials Science and Engineering, University of California Berkeley, Berkeley, California 94720, United States}
\affiliation{Materials Science Division, Lawrence Berkeley National Laboratory, Berkeley, California 94720, United States}

\author{Emily\,C.\,Schueller}
\affiliation{Materials Research Laboratory, University of California, Santa Barbara, California 93106, United States}
\affiliation{Materials Department, University of California, Santa Barbara, California 93106, United States}

\author{Kristin\,A.\,Persson}
\affiliation{Department of Materials Science and Engineering, University of California Berkeley, Berkeley, California 94720, United States}
\affiliation{Molecular Foundry, Lawrence Berkeley National Laboratory, Berkeley, California 94720, United States}

\author{Ram\,Seshadri}
\affiliation{Materials Research Laboratory, University of California, Santa Barbara, California 93106, United States}
\affiliation{Materials Department, University of California, Santa Barbara, California 93106, United States}
\affiliation{Department of Chemistry and Biochemistry, University of California, Santa Barbara, California 93106, United States}

\title[Magnetism and magneticaloric]{Magnetism and magnetocaloric properties of Co$_{1-x}$Mn$_x$Cr$_2$O$_4$}
\sloppy

\begin{abstract}
Co$_{1-x}$Mn$_x$Cr$_2$O$_4$ crystallizes as a normal spinel in the cubic $Fd \overline{3}m$ space group, and the end members have been 
reported to display a region of collinear ferrimagnetism as well as a low-temperature spin-spiral state with variable coherence lengths 
from 3\,nm to 10\,nm in polycrystalline samples. Here, we present the synthesis of the entire solid solution, and data showing that the 
ferrimagnetic ordering temperature as well as the spin-spiral lock-in temperature are tunable with the Co/Mn ratio. The peak magnetocaloric 
entropy change was determined to be $\Delta S_M$\,=\,$-$5.63 \,J\,kg\,$^{-1}$K\,$^{-1}$ in an applied magnetic field change of 
$\Delta H$\,=\,0\,T to 5\,T for the Mn end-member at the ferrimagnetic ordering temperature. 
Using density functional theory (DFT), we explore the shortcomings of the magnetic deformation proxy to identify trends in 
$\Delta S_M$ across composition in this spinel system, and explore future extensions of theory to address these discrepancies. 
\end{abstract}

\maketitle
\section*{Introduction}
In the search for alternative refrigeration technologies, magnetic refrigeration has emerged as an environmentally friendly and more efficient alternative to conventional vapor-compression refrigeration\,\cite{Franco2012}. Magnetic refrigeration uses the magnetocaloric effect (MCE) by alternating between states of high and low entropy (low and high magnetization, respectively) to impart a reversible temperature change.
Since the discovery of the giant magnetocaloric effect in Gd$_5$Si$_2$Ge$_2$\,\cite{Pecharsky1997}, research has focused on methods of understanding which materials may exhibit a large magnetocaloric effect. Saturation magnetization (\MSexp{}) is typically used as an indicator for the magnitude of the magnetocaloric effect, and high magnetization compounds have been thought to be well-performing magnetocalorics. 

Recent analysis has shown \MSexp{} is not always the most effective predictor of which materials may exhibit favorable magnetocaloric properties, and \MSexp{} shows poor correlation with the MCE as quantified by the magnetic entropy change, $\Delta S_M$\,\cite{Bocarsly2017}. As such, there has been an increase in interest in systems whose lattice and spin degrees of freedom are strongly coupled, especially in magnetocaloric materials. It has been proposed that this coupling of spin and lattice -- termed magnetostructural coupling -- is important to understanding, and even predicting, the magnitude of the magnetocaloric effect in ferromagnets\,\cite{Bocarsly2017}. 

For many decades, the MCE has been used to reach cryogenic (mK and $\mu$K) temperatures\,\cite{Giauque1933,Hamilton2014,Mukherjee2017}. Recently, there has been great interest in magnetocalorics for room-temperature magnetic refrigeration\,\cite{Gschneidner2008}. This necessitates magnetic ordering temperatures near room temperature such that the maximum $\Delta S_M$ (and, therefore, MCE)t will be near room temperature. However, other technologies exist that would require excellent magnetocaloric performance in more intermediate temperature ranges below the boiling point of liquid nitrogen (\textit{e.g.} 2\,K to 70\,K). This includes staged/cascaded cycles necessary to cover the wide temperature range required for hydrogen liquefaction\,\cite{Iwasaki2003}, which boils near 20\,K. While some effort has been made toward studying inexpensive materials like chalcogenide spinels\,\cite{Bouhbou2017}, currently much of the research into magnetocalorics viable for hydrogen liquefaction involves expensive rare-earth elements\,\cite{Nakagawa2006,Luo2007,Zhu2011}, stemming partially from the fact that these often exhibit high \MSexp{} values.  Additionally, a segmented series of materials with gradually changing transition temperatures would be beneficial to reach hydrogen liquefaction temperatures\,\cite{Numazawa2014}. Here, we look to an earth-abundant transition metal oxide solid solution -- Co$_{1-x}$Mn$_x$Cr$_2$O$_4$ -- with gradually varying transition temperatures magnets as a more cost-effective alternative to rare-earth magnetic materials.

We investigate how compositional changes and slight structural changes affect the magnetism and MCE as quantified by the magnetic entropy change $\Delta S_M$ in the spinel solid solution Co$_{1-x}$Mn$_x$Cr$_2$O$_4$. This series of compounds is geometrically frustrated stemming partially from the 3D-pyrochlore sublattice of the Cr$^{3+}$ and partially from the diamond sublattice formed by cations in the Mn/Co site. Frustrated spin systems often possess interesting and exotic ground states\,\cite{Ramirez1994, Kimura2007}, such as the spiral spin textures found in the Mn\,\cite{Tobia2015} and Co\,\cite{Lawes2006,Chang2009} chromite spinels, and have been linked to good magnetocaloric performance \cite{Zhitomirsky2003}.  According to a 2014 study by Dey et al., MnCr$_2$O$_4$ exhibits some degree of magnetostructural coupling and shows magnetoelastic transitions at both the N\'{e}el temperature and the spin spiral lock-in temperature \cite{Dey2014}, and its geometric frustration likely plays a part in coupling the spin and lattice\,\cite{Moessner2006}. As the first study of the magnetocaloric properties of this solid solution, we find that the Mn-rich members exhibit a high $\Delta S_M$ of up to $-$5.67\,J\,kg\,$^{-1}$K\,$^{-1}$ for a field change of 0\,T to 5\,T.

\section{Materials and Methods}
\subsection{Synthesis}
Polycrystalline powders of the solid solution series Co$_{1-x}$Mn$_x$Cr$_2$O$_4$ were prepared using solid-state synthesis. Precursor materials CoC$_2$O$_4$ $\bullet$ 2H$_2$O, MnO, and Cr$_2$O$_3$ were used as received.  For clarity, herein samples will be referred to according to their nominal $x$ values (i.e. $x$\,=\,0.00, 0.25, 0.50, 0.75, and 1.00).  CoCr$_2$O$_4$ was synthesized following literature procedure \cite{Lawes2006} but regrinding and annealing was not found to improve phase purity so samples were heated at 800 \textdegree C for 24\,hours, then at 1000 \textdegree C for 24\,hours without removing from the furnace, then the furnace was allowed to cool. Samples $x$\,=\,0.25 and 0.50 were also synthesized using this procedure. However, due to initial phase separation into two different spinel phases with slightly different unit cell parameters, samples $x$\,=\,0.25 and 0.50 were quenched from 1000 \textdegree C. MnCr$_2$O$_4$ was also synthesized according to literature \cite{Winkler2009}, and $x$\,=\,0.75 also followed this procedure.

\subsection{Structural characterization}

High-resolution synchrotron powder X-ray diffraction (XRD) data were collected at room temperature on all samples using beamline 11-BM at the Advanced Photon Source (APS), Argonne National Laboratory. For room temperature scans, powderized samples were loaded into 0.8\,mm diameter Kapton capillaries with each end sealed with clay and measured for 6.7 minute scans. $x$\,=\,0.00, 0.75, and 1.00 were measured with $\lambda$\,=\,0.457850\,\AA\ and $x$\,=\,0.25 and 0.50 were measured at another time with $\lambda$\,=\,0.457845\,\AA. Rietveld refinements of data were performed using TOPAS \cite{Coelho2018}. All samples were fit with size and strain parameters, except the complex peak shape of $x$\,=\,0.25 which required a 2$\theta$-dependent split Pearson VII peak shape to describe variation in peak asymmetry fully. Structures were visualized using VESTA-3 \cite{Momma2011}.

\subsection{Magnetic measurements}
Magnetic properties were measured on 5\,mg to 10\,mg of powder loaded into capillaries and measured a Quantum Design MPMS3 equipped with a vibrating sample magnetometer (VSM). Zero field-- and field--cooled magnetization ($M$) vs temperature ($T$) measurements were taken upon warming at a rate of 5\,K\,min$^{-1}$. In order to determine $\Delta S_M$, $M$ versus $T$ measurements were taking on cooling (using a rate of 5 \,K\,min$^{-1}$) at various fields from $H$\,=\,0.1\,T to $H$\,=\,5\,T. Temperature derivatives of the $M(T)$s were calculated using Tikhonov regularization \cite{Stickel2010}, and then integrals with field were calculated using the trapezoid method to obtain $\Delta S_M$. Data were analyzed using the \texttt{magentro.py} code, and more details of this procedure have previously been reported \cite{Bocarsly2018}.

\subsection{First-principles calculations}
The crystal structures corresponding to the $x=0$ and $x=1$ endpoints of the spinel system Mn$_x$Co$_{1-x}$Cr$_2$O$_4$ (CoCr$_2$O$_4$ and MnCr$_2$O$_4$, respectively) were obtained from the Materials Project database\cite{Jain2018}. To identify the energetically stable collinear magnetic ordering of these spinel structures, an enumeration of the possible magnetic orderings and evaluation of their respective energies was performed using the \textit{atomate} workflow \cite{horton_high-throughput_2019}. 

Despite the experimental results in support of both CoCr$_2$O$_4$ and MnCr$_2$O$_4$ exhibiting a spin-spiral ground state \cite{yoon_nmr_2010, PhysRevB.70.214434, Dey2014, Lin2017, Lawes2006, Yang2012}, we performed collinear ferrimagnetic calculations. The justification for this approach is based on experiments that show that the spin-spiral transition temperature is significantly lower than the N\'{e}el temperature, $T_N$. Furthermore, we are interested in the ferrimagnetic (FiM) to paramagnetic (PM) phase transition, which yields the greatest change in entropy. Based on the experimentally reported values of the spin-spiral transition temperatures for CoCr$_2$O$_4$ and MnCr$_2$O$_4$, the order-disorder phase transition has been experimentally determined to involve a predominantly collinear FiM to PM transition.

Using the endpoint spinel structures, a set of possible structures of each composition Mn$_x$Co$_{1-x}$Cr$_2$O$_4$ were enumerated based on their crystallographic symmetries, up to a supercell size of two times the formula unit using the \textit{pymatgen} interface with \textit{enumlib} \cite{pymatgen, enumlib2008, enumlib2009, enumlib2012}.

Following the generation of intermediate crystal structures for $x = $ 0.25, 0.5, and 0.75, we generalized the collinear FiM ordering that would be expected for the spinel system endpoints, with a net spin-up moment on the Cr atoms, and spin-down moment on the Mn and Co atoms. This collinear FiM ordering was used in spin-polarized calculations for computing the deformation proxy, $\Sigma_M$, and saturation magnetization, \MScomp{}.

\section{Results and Discussion}
\subsection{Structure and Phase Purity}
\begin{figure}
\includegraphics[width=0.45\textwidth]{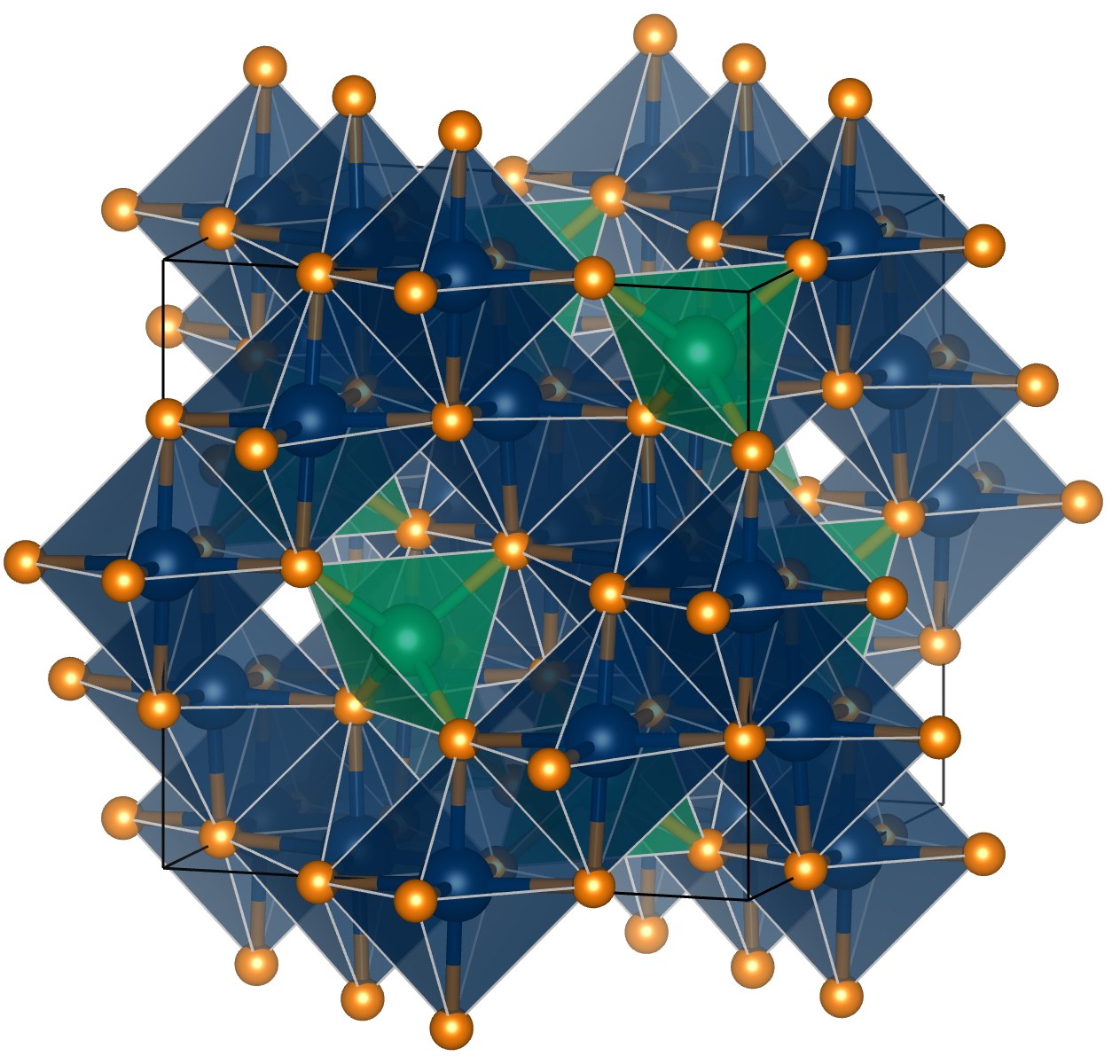}
\caption{A view of one unit cell of the spinel structure type, $A$Cr$_2$O$_4$ ($A$ = Co, Mn) where the $A$ site is tetrahedrally coordinated by O and corner shared, and the Cr site is octahedrally coordinated and edge- and corner shared.}
\label{Fig:Structure}
\end{figure}

The spinel structure type is pictured in Figure\,\ref{Fig:Structure} where Co/Mn atoms are in green corner-shared $A$-site tetrahedra, Cr atoms are in deep blue edge-and corner-shared $B$-site octahedra, and O atoms are in tangerine at the vertices of the tetrahedra and octahedra. The normal spinel structure type exists such that divalent cations are in tetrahedral $A$-sites and trivalent cations are in octahedral $B$-sites.  Many spinel compositions have the ability to form partially- or fully-inverted spinels such that $A$-site atoms oxidize to trivalent and occupy the $B$-site, and $B$-site atoms reduce to divalent and occupy the $A$-site. Chromite spinels, however, are a unique case. Since trivalent Cr is $d^3$, it is more stable in an octahedral crystal field splitting arrangement, where all spins are in the lowest energy, triply-degenerate $t_{2g}$ orbitals. Trivalent Cr in a tetrahedral crystal field would necessarily have two low-energy spins in the doubly degenerate $e_g$ orbital and one high energy spin in the $t_{2g}$ orbitals, destabilizing this arrangement relative to the octahedral arrangement. Thus, in chromite spinels we can be relatively certain that Co and Mn only occupy the tetrahedral $A$-sites while Cr only occupies the octahedral $B$-site.

\begin{figure}
\includegraphics[width=0.45\textwidth]{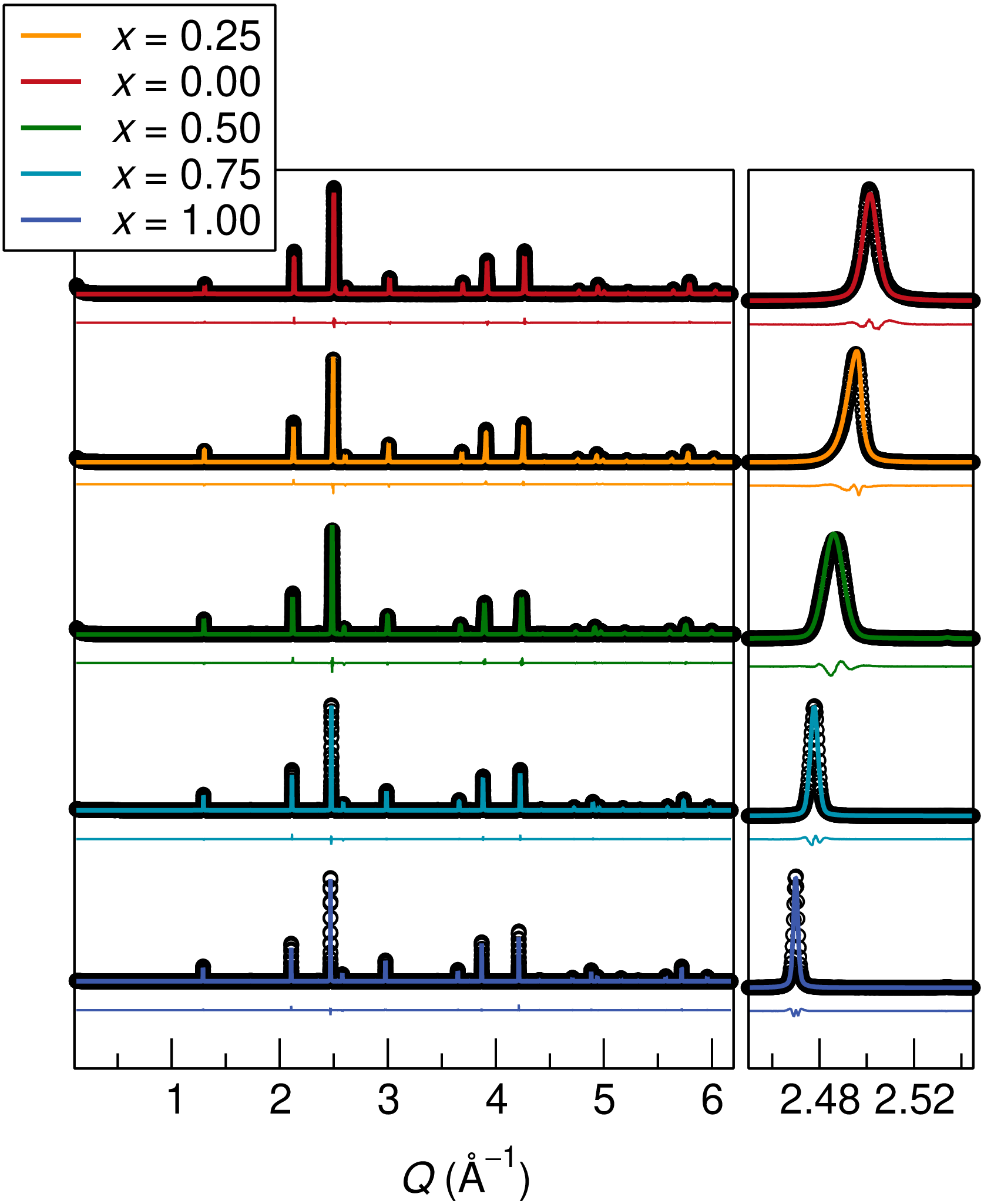}
\caption{The results of Rietveld refinement of synchrotron X-ray diffraction data measured for each sample show the spinel structure type. The data are black circles, the overlaying line represents the model based on the structure, and the line below each pattern indicates the difference between the data and model. A close view of the (311) peak (right) indicates a complete solid solution at the resolution of the instrument.}
\label{Fig:XRD}
\end{figure}

Rietveld refinement results of synchrotron powder x-ray diffraction for samples in this study are shown in Figure\,\ref{Fig:XRD}. The left panel shows that each sample fits to a single spinel phase at the resolution of the instrument. The right panel shows a close view of the evolution in peak position of the (311) peak as a function of $x$: a monotonic decrease in $Q$ corresponding to an increase in unit cell parameter.  We note that samples with $x$\,=\,0.25 and $x$\,=\,0.50 show slightly more broadening than other samples. These samples did require quenching from maximum temperature in order to yield single phase samples, and we note that there could be two spinel phases with slightly different unit cell parameters, yet even the extremely high resolution of beamline 11-BM at the Advanced Photon Source ($<$1.4 $\times$ 10$^{-4} \Delta Q/Q$) is not able to resolve two different phases, so we proceed as if each sample is single phase. There are no crystalline ferromagnetic impurities present in the samples, however $x$\,=\,0.50, 0.75, and 1.00 contain a small ($<$1.5\,wt\%) amount of unreacted Cr$_2$O$_3$, which is antiferromagnetic at room temperature \cite{Corliss1965}, that could not be eliminated by further reaction.

\begin{figure}
\includegraphics[width=0.45\textwidth]{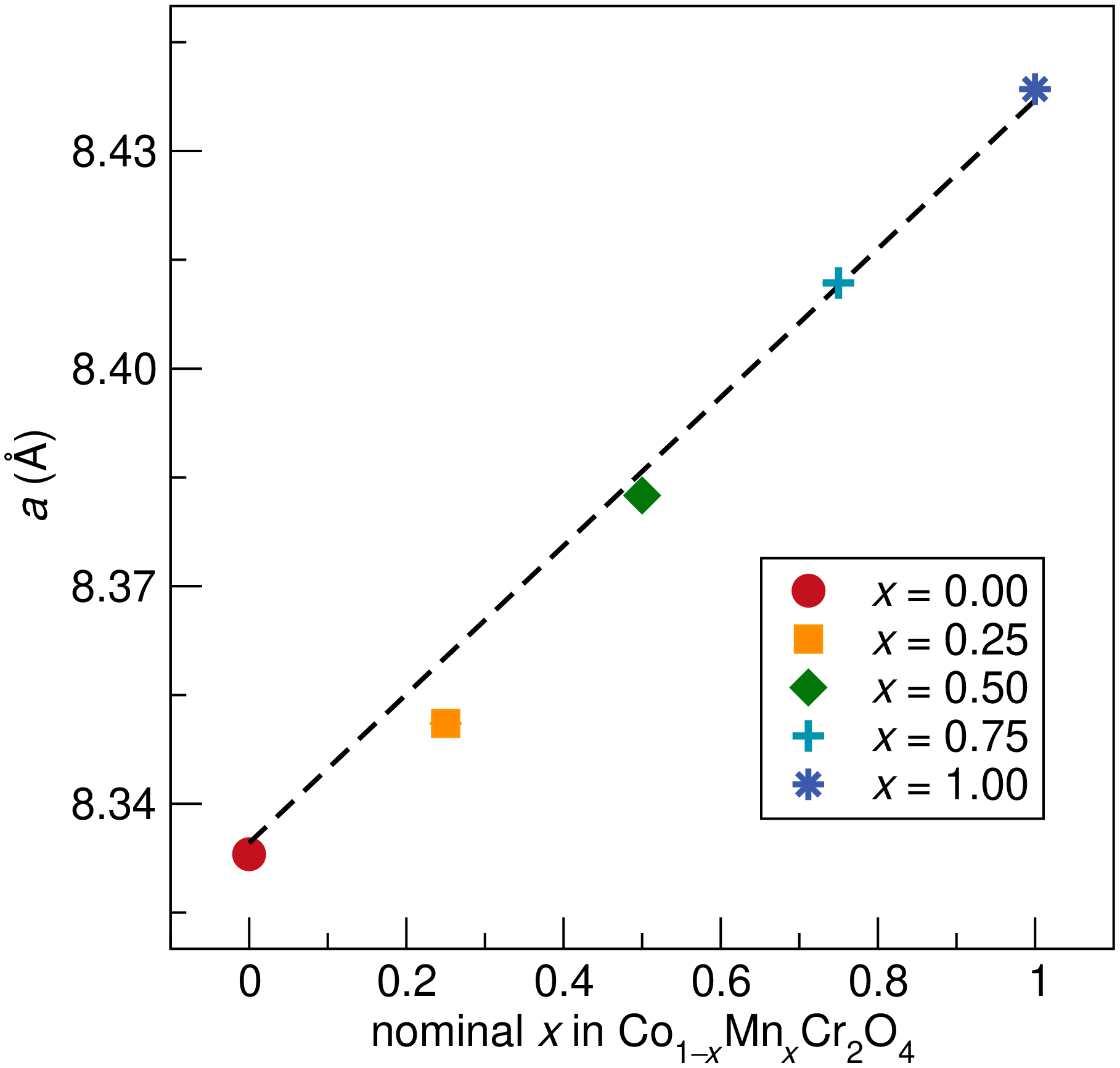}
\caption{a) Lattice parameters from Rietveld refinement of synchrotron X-ray diffraction data for each sample show that each follows the V\'{e}gard law (dashed line) as expected for a solid solution.}
\label{Fig:LPXRF}
\end{figure}

Further results of Rietveld refinements, including the weighted profile R-value $R_{wp}$, are tabulated Table\,\ref{Table:Rietveld}. The cell parameter $a$ (also depicted in Figure\,\ref{Fig:LPXRF}) and oxygen position $u$ were determined during Rietveld refinement. The ionic radii of tetrahedrally coordinated Co$^{2+}$ and Mn$^{2+}$ are 0.72\,\AA\ and 0.80\,\AA\ \cite{Shannon1976}, respectively, so it follows that increasing Mn composition should increase the lattice parameter throughout the solid solution. The anion in spinel structures -- here, O -- is located on the crystallographic equipoint $32e$ and variation in this position, represented as the value $u$, reflects how the structure changes when accommodating different sizes of cations. When $u$\,=\,0.25, the anions are in ideal cubic closest-packing arrangement with perfect CrO$_6$ octahedra; increases in $u$ above the ideal value of 0.25 indicate the size of the tetrahedron is increasing, and the octahedron is shrinking and undergoing a trigonal compression \cite{Hill1979}. Table\,\ref{Table:Rietveld} shows that the $u$ parameter overall increases with increasing Mn substitution, shifting it higher than the ideal value, and implying the tetrahedral site size increases across the solid solution, in good agreement with the unit cell parameter increase and larger ionic radius of Mn$^{2+}$.

\begin{table}
\caption{Crystallographic Data for Co$_{1-x}$Mn$_x$Cr$_2$O$_4$.}
\begin{tabular}{@{}lllll@{}}
\botrule
 x & $a$ (\AA) & $u$  &  $R_{wp}$ & Cr$_2$O$_3$ (\%)  \\ 
 \hline
 0.00 & 8.33304(4) & 0.26125(9) & 14.46  & --  \\
 0.25 & 8.34771(4) & 0.25949(1) & 10.81 & --  \\
 0.50 & 8.3851(4)   & 0.26043(6)   & 12.149 & 0.81(1)   \\
 0.75 & 8.41180(2) & 0.26241(7) & 13.72 & 0.54(1)  \\
 1.00 & 8.43853(1) & 0.26385(6) & 12.82 & 1.40(2) \\ \botrule
\end{tabular}
\label{Table:Rietveld}
\end{table}

Figure\,\ref{Fig:LPXRF} shows that the lattice parameter $a$ of the series increases with increasing $x$ and follows V\'{e}gard's law as indicated by the dashed line (created from the literature lattice parameters of the end members CoCr$_2$O$_4$ \cite{Casado1986} and MnCr$_2$O$_4$ \cite{Hastings1962}). This is in good agreement with the larger ionic radius of tetrahedral Mn$^{2+}$ as compared to Co$^{2+}$. Each data point is also in good agreement with the V\'{e}gard line indicating that this series forms a complete solid solution.

\subsection{Magnetic Properties}
\begin{figure}
\includegraphics[width=0.45\textwidth]{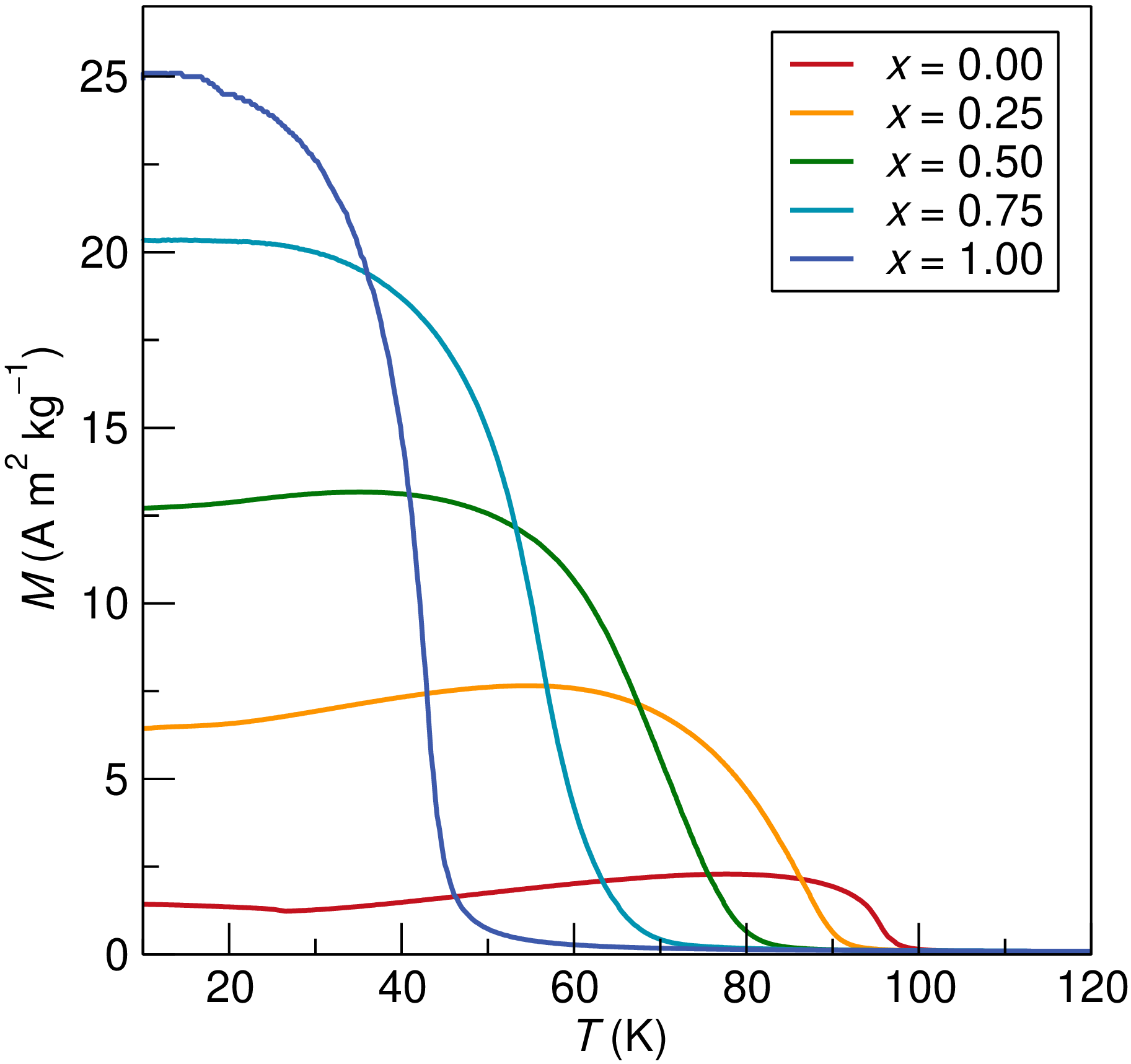}
\caption{Temperature-dependent magnetization data reveal an increase in magnetization and decrease in ordering temperature as Mn content is increased.}
\label{Fig:MTs}
\end{figure}

Both the Co and Mn end members of the solid solution undergo a paramagnetic to ferrimagnetic transition and, after a region of collinear ferrimagnetic order, have a complex spiral ground state at low temperatures. In Figure\,\ref{Fig:MTs}, which shows temperature-dependent magnetization for each composition, the transition from the paramagnetic state to the collinear ferrimagnetic state can be seen to decrease as the Mn composition is increased. This is important from an engineering standpoint as tunability and control of transition temperature, and thus peak $\Delta S_M$ (discussed later), in this range is especially valuable to cascaded magnetic refrigeration for hydrogen liquefaction.

The high temperature (300\,K to 390\,K) inverse susceptibility of each sample was fit to a linear regression and the equation of the resulting line was used to extract parameters from the Curie-Weiss equation, $\chi$\,=\,$C$/($\chi - \Theta_{CW}$) -- namely, the effective paramagnetic moment \MUeff{}, Curie constant $C$, and Curie-Weiss intercept $\Theta_{CW}$. These values are presented in Table\,\ref{Table:Mu_eff}. The spin-only formula for  \MUeff{} is generally taken as valid for 3$d$ transition metal ions, however it only strictly applies to specific cases. In reality, the orbital contribution is not totally quenched and spin-orbit coupling plays a role in determining \MUeff{}. $d^1-d^4$ transition metal ions have a spin orbit coupling constant smaller than zero, and their  \MUeff{} values tend to be smaller than what is calculated by the spin-only formula. $d^6-d^9$ transition metals have a spin-orbit coupling constant larger than zero and tend to have  \MUeff{} larger than calculated by spin-only formula. The samples in this series show a stochastic trend with composition concerning how they compare to the spin-only and unquenched \MUeff{} values. Co$^{2+}$ is $d^7$ and we would expect compounds rich in Co$^{2+}$ to deviate higher than the spin-only value; conversely Mn$^{2+}$ is $d^4$ and we would expect that Mn-rich samples deviate lower than the spin-only value. Values of \MUeff{} from the Co- and Mn-rich ends of the solid solution are both far above unquenched estimates and far below spin-only estimates with no real trend to speak of. This may be due to not fitting completely in the paramagnetic regime for each of these samples as, due to instrumentation and sample limits, only measurement to 390\,K is practical. As $x$ increases, the magnitude of $\Theta_{CW}$ decreases suggesting that the number of dominant antiferromagnetic Co/Mn interactions with Cr decreases as Mn content is increased, and the decrease is overall monotonic.

\begin{widetext}

\centering

\begin{table}[h]
\caption{Results from fitting inverse magnetic susceptibility data of Co$_{1-x}$Mn$_x$Cr$_2$O$_4$ to a linear regression to extract parameters in the Curie-Weiss equation. $T_C$ is assumed to be the peak of the $\Delta S_M$ curve. Estimated moments (spin-only and unquenched) are calculated using: $ \mueff{} = \sqrt{2\mu_{\text{Cr}}^2+(1-x)(\mu_{\text{Co}})^2+x(\mu_{\text{Mn}})^2}$. Individual moments, such as $\mu_{\text{Mn}}$ are calculated as $\mueff = g\sqrt{S(S+1)}$ using an isotropic Land\'e g factor of 2.} 

\label{Table:Mu_eff}
\begin{tabular}{ccccccc}
\botrule
& & \multicolumn{3}{c}{\MUeff{} ($\mu_B$)} &   \\ \cline{3-5}
$x$ & \textit{C} ( [$\mu_B$ / f.u.] $T^{-1}$ K ) & measured & spin-only & unquenched & $\Theta_{CW}$ (K) & $T_C$(K)  \\
0.00 & 7.83 & 7.9 & 6.7 & 7.5 & $-$613 & 97.9 \\
0.25 & 7.00 & 7.5 & 7.1 & 7.7 & $-$471 & 88.7 \\
0.50 & 6.96 & 7.5 & 7.4 & 7.8 & $-$369 & 74.2 \\
0.75 & 7.82 & 7.9 & 7.7 & 7.9 & $-$337 & 58.8 \\
1.00 & 7.53 & 7.8 & 8.1 & 8.1 & $-$271 & 43.7 \\ \botrule
\end{tabular}
\end{table}

\begin{table}[h]
\caption{Results from fitting inverse magnetic susceptibility data of Co$_{1-x}$Mn$_x$Cr$_2$O$_4$ to Equation \ref{eq:cw_general} using \CurveFitName{}.  $T_C$ values are reproduced from Table \ref{Table:Mu_eff}. We hypothesize that the ``$A$'' sublattice corresponds to tetrahedral sites in the spinel structure occupied by Mn and Co species, and the ``$B$'' sublattice corresponds to octahedral sites occupied by Cr atoms.} 

\begin{tabular}{c|cc|ccc|ccc}
\botrule

$x$	&	$|\Theta_{CW}|$	&	$T_C$	&	$C_B$			&	$C_A$			&	$C_A + C_B$	&	$\lambda_{BB}$			&	$\lambda_{AA}$			&	$\lambda_{AB}$			\\
	&	\multicolumn{2}{c}{(K)}			&	\multicolumn{3}{|c}{	( [$\mu_B$ / f.u.] T$^{-1}$ K )	}							&	\multicolumn{3}{|c}{	( [$\mu_B$ / f.u.]$^{-1}$ T )	}									\\
\hline																											
0.00	&	97.4	&	97.9	&	0.95	$\pm$	0.01	&	9.53	$\pm$	0.11	&	10.48	&	66.5	$\pm$	0.2	&	86.2	$\pm$	0.4	&	127.6	$\pm$	0.6	\\
0.25	&	89.6	&	88.7	&	0.89	$\pm$	0.01	&	7.85	$\pm$	0.07	&	8.74	&	67.1	$\pm$	0.3	&	72.0	$\pm$	0.4	&	118.4	$\pm$	0.6	\\
0.50	&	78.7	&	74.2	&	6.76	$\pm$	0.02	&	1.20	$\pm$	0.00	&	7.96	&	71.7	$\pm$	0.1	&	-37.7	$\pm$	0.0	&	48.2	$\pm$	0.0	\\
0.75	&	65.8	&	58.8	&	8.66	$\pm$	0.03	&	0.57	$\pm$	0.00	&	9.23	&	60.2	$\pm$	0.1	&	-113.0	$\pm$	0.0	&	15.1	$\pm$	0.1	\\
1.00	&	45.4	&	43.7	&	8.37	$\pm$	0.11	&	2.54	$\pm$	0.01	&	10.91	&	73.6	$\pm$	0.4	&	1.3	$\pm$	0.1	&	38.9	$\pm$	0.0	\\

\botrule
\end{tabular}
\label{Table:CWsublattice_params}
\end{table}

\end{widetext}

\begin{figure}
\includegraphics[width=0.45\textwidth]{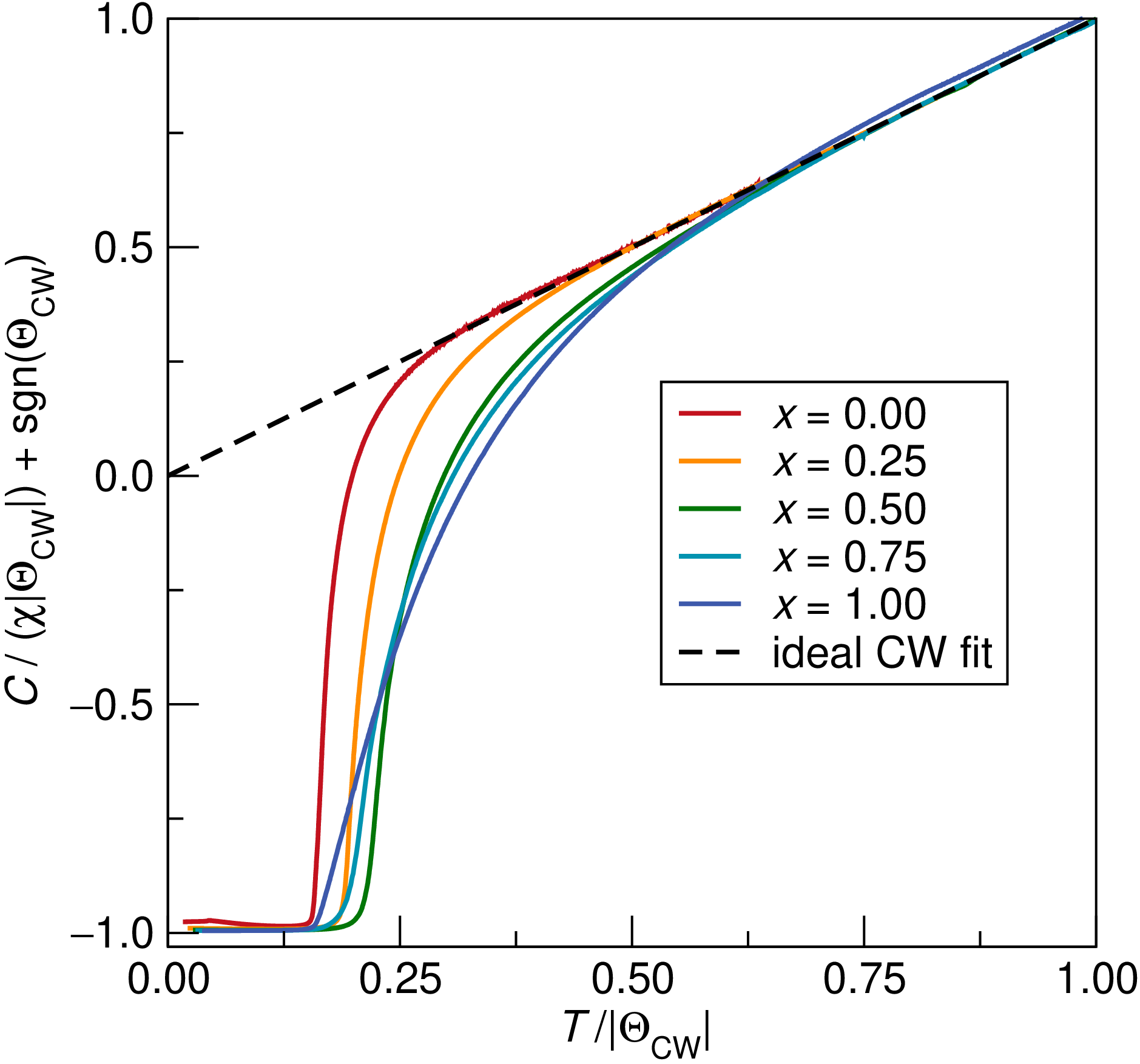}
\caption{Scaled inverse susceptibility data as a function of scaled temperature as described by equation are shown. The dashed line indicates simplified (ferromagnetic) Curie-Weiss paramagnetism. Negative deviations in all samples reflect uncompensated interactions and suggest ferrimagnetic interactions.}
\label{Fig:CWs}
\end{figure}

\begin{figure}
\includegraphics[width=0.45\textwidth]{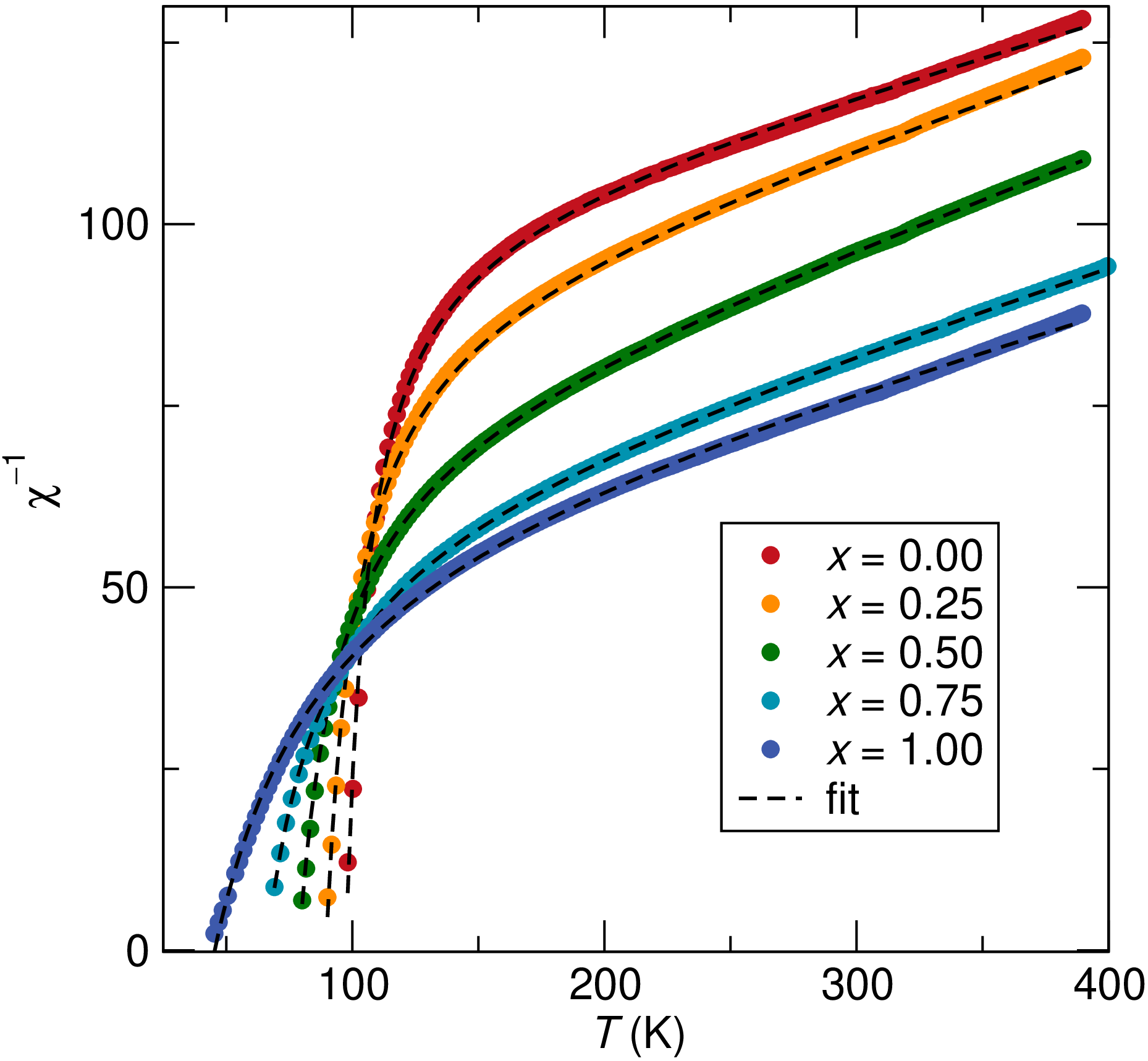}
\caption{The inverse susceptibility is plotted versus temperature for each composition, $x$ in Co$_{1-x}$Mn$_x$Cr$_2$O$_4$. The lighter markers indicate the model equation fit to the data, Equation \ref{eq:cw_general}, for two magnetic sublattices. The fitted parameters of this equation are supplied in Table \ref{Table:CWsublattice_params}.}
\label{Fig:CWsublattice}
\end{figure}

A plot of $C/(\chi|\Theta_{CW}|)+sgn(\Theta_{CW})$ vs $T/\Theta_{CW}$ collapses all high temperature susceptibility data as shown in Figure\,\ref{Fig:CWs}. The dashed straight line intersecting the origin corresponds to ideal Curie-Weiss behavior; since the high temperature data fit well onto this line, this indicates that the fits in the high temperature regime are valid. Plotting the susceptibility data in this manner allows us to understand the nature of dominant magnetic exchange interactions \cite{Melot2009,Neilson2011}. Since all show negative deviations from the ideal Curie-Weiss line, this indicates uncompensated antiferromagnetism manifesting as ferrimagnetism.

From inspecting Table \ref{Table:Mu_eff}, it is clear that $|\Theta_{CW}|$ severely overestimates $T_C$. This is not entirely surprising, because this disagreement is expected for more than one anti-parallel sublattice \cite{kittel_introduction_2005}. We explore the extension of Weiss mean field theory to two magnetic sublattices in Section \ref{section:cw_sublattice_appendix}. By fitting Equation \ref{eq:cw_general} to $1/\chi$ data versus temperature, we were able to achieve agreement between $|\Theta_{CW}|$ and $T_C$ to within a few Kelvin, as reported in Table \ref{Table:CWsublattice_params}. This Curie-Weiss temperature, $\Theta_{CW}$, is defined as $|\Theta_{CW}| = \max\left\{ \eta_i \left[ -\bm W \right] \right\}$, as explained in Section \ref{section:cw_sublattice_appendix}. 

The effectiveness of this theoretical extension is exemplified in the agreement of the model to $1/\chi$ at all temperatures greater than $T_C$, as shown in Figure \ref{Fig:CWsublattice}. The parameters of Equation \ref{eq:cw_general} are reported in Table \ref{Table:CWsublattice_params}, with their associated uncertainty values output by \CurveFitName{}. The ``$A$" and ``$B$" labels indicate a separate magnetic sublattice. Based on the anticipated low temperature ferrimagnetic ordering of this spinel system, these lattices should each correspond to either tetrahedral sites, occupied by Mn or Co atoms, or octahedral sites, occupied by Cr atoms. Observing that $\lambda_{BB}$ remains relatively constant versus Mn composition $x$ compared to the other values of $\lambda$, we anticipate that the $B$-sublattice corresponds to the octahedral (Cr) sublattice, and the $A$-sublattice to the tetrahedral (Mn, Co) sublattice.

\begin{figure}
\includegraphics[width=0.45\textwidth]{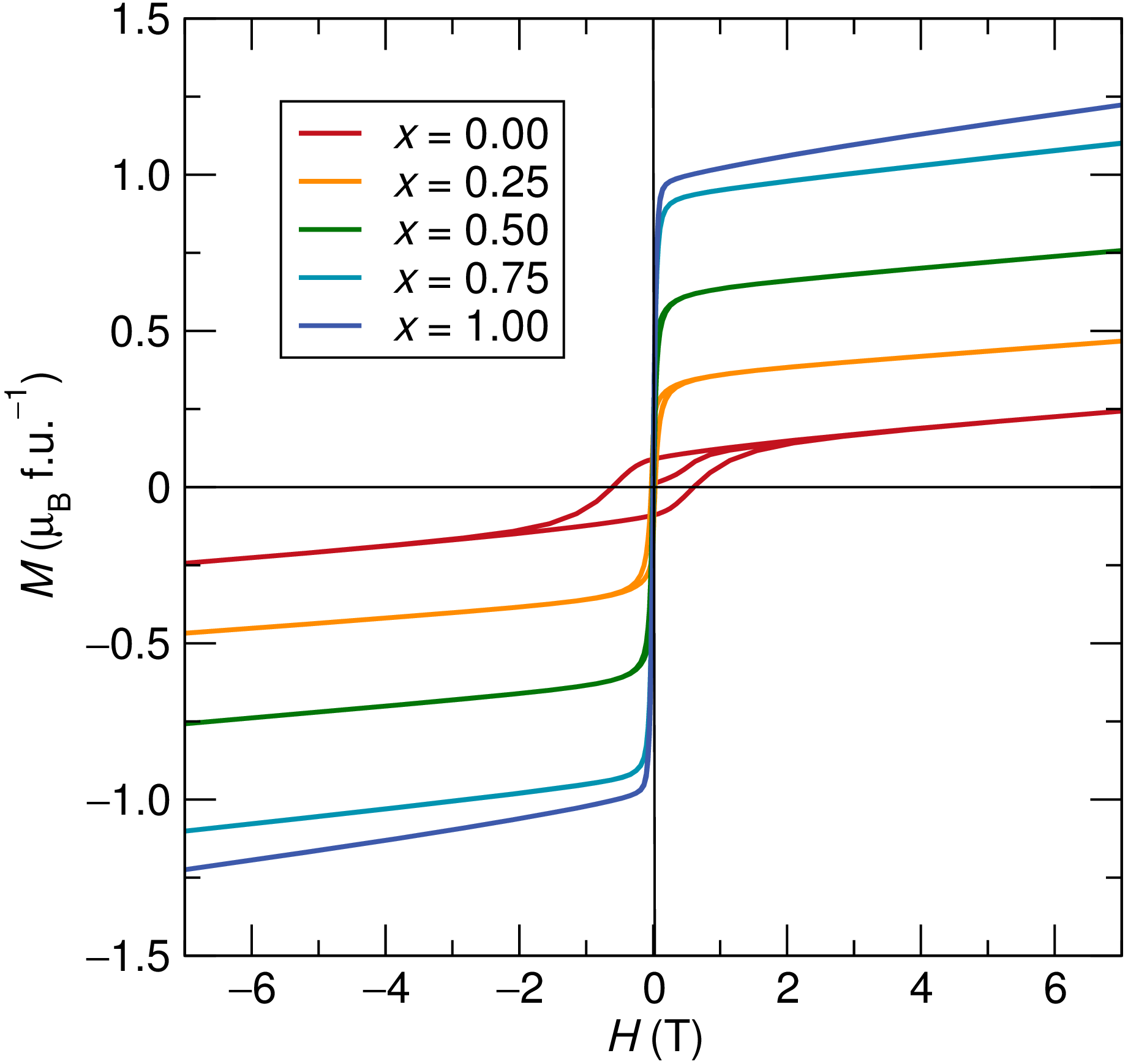}
\caption{Field-dependent isothermal magnetization of Co$_{1-x}$Mn$_x$Cr$_2$O$_4$ were taken at 30 K for all samples to avoid the spiral magnetic ordering region and reflect the magnetization in the collinear ferrimagnetic regime. Increasing Mn content increases the saturation magnetization.}
\label{Fig:MHs}
\end{figure}

\begin{table}
\begin{threeparttable}
\caption{Spontaneous (\Mspont{}) and Saturation Magnetization (\MSexp{}) Data for Co$_{1-x}$Mn$_x$Cr$_2$O$_4$.}
\label{Table:Sat}
\begin{tabular}{@{}cccc@{}}
\botrule
 $x$ & \Mspont{} ($\mu_B$/f.u.)$^a$ & \MSexp{} ($\mu_B$/f.u., 2 T) &  \MSexp{} ($\mu_B$/f.u., 5 T)  \\ 
 \hline
 0.00 & --    & 0.14 & 0.21   \\
 0.25 & 0.32 & 0.38 & 0.44   \\
 0.50 & 0.60 & 0.68   & 0.72    \\
 0.75 & 0.88 & 0.92 & 0.99   \\
 1.00 & 0.97 & 1.06 & 1.17  \\ \botrule
\end{tabular}
\begin{tablenotes}
\item $^a$Extrapolated values are from 30\,K isotherms for each sample
\end{tablenotes}
\end{threeparttable}

\end{table}

Figure\,\ref{Fig:MHs} shows field-dependent magnetization data of each sample at 30\,K. This temperature was chosen to avoid the spiral magnetic ordering region and reflect the magnetization of in the collinear ferrimagnetic region in each sample. The $x$\,=\,0.00 end has considerable coercivity as compared to the other compositions, and this coercivity has been seen to increase with decreasing temperature (from 75\,K to 3\,K)\,\cite{Mohanty2018}. Since none of the samples fully saturates at the highest applied field of 7\,T, Table\,\ref{Table:Sat} shows spontaneous magnetization (\Mspont{}) as extrapolated from Arrott-Belov plots (which could not be determined for $x$\,=\,0.00) as well as saturation magnetization (\MSexp{}) values at fields of 2\,T and 5\,T. Overall, this table shows that magnetization values increase as Mn content is increased. Also, none of the individual magnetization values are very large, again indicating that \MSexp{} is not always a useful indicator of magnetocaloric viability, discussed further below.

\subsection{Magnetocaloric Properties}

\begin{figure}
\includegraphics[width=0.45\textwidth]{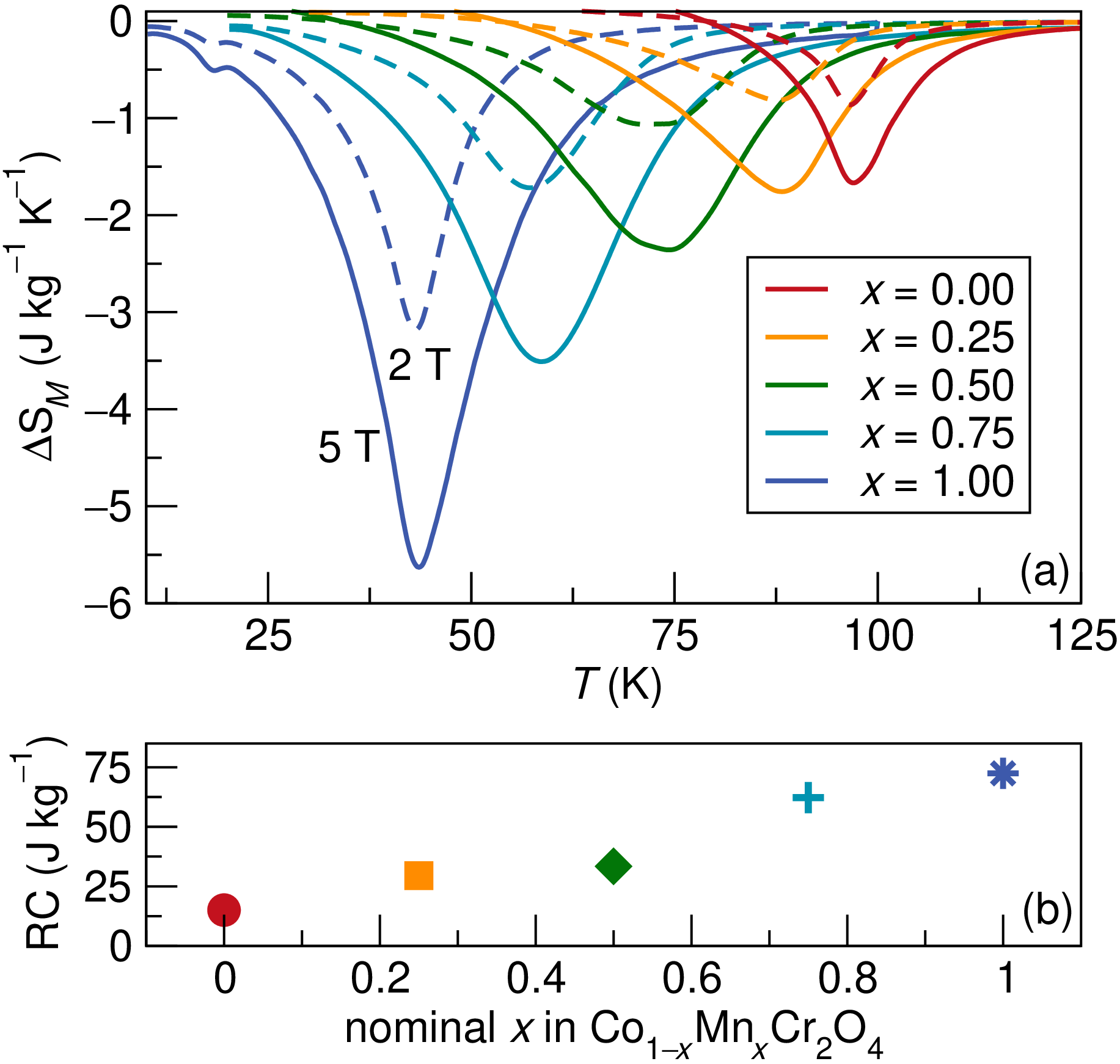}
\caption{(a) Calculated $\Delta S_M$ curves for each sample for field changes of 0\,T to 2\,T (dashed lines) and 0\,T to 5\,T (solid lines) in the species shows an increasing peak $\Delta S_M$ and decreasing peak temperature with increasing Mn content. (b) Increasing RC with increasing Mn content indicates that the overall performance of the magnetocaloric increases with Mn content.}
\label{Fig:DSMRC}
\end{figure}

The magnitude of the magnetocaloric effect can be derived from temperature-dependent magnetization measurements at varying fields which are then derived and integrated over field to yield $\Delta S_M$, depicted in Figure\,\ref{Fig:DSMRC}(a). The absolute value of $\Delta S_M$ increases as Mn content increases, and the temperature occurrence of the peak value decreases as expected from previously discussed temperature-dependent magnetization measurements. This series spans a wide range of peak $\Delta S_M$ values, from $-$1.67\,J\,kg\,$^{-1}$K\,$^{-1}$ for $x$\,=\,0.00 to $-$5.63\,J\,kg\,$^{-1}$K\,$^{-1}$for $x$\,=\,1.00. It is difficult to visually determine whether or not $\Delta S_M$ peaks broaden or sharpen throughout the series, and this can affect overall magnetocaloric performance. Thus we employ a more mathematically rigorous method of understanding magnetocaloric performance, the refrigerant capacity. This involves integrating the area under the $\Delta S_M$ curve using the full width at half maximum of each curve as temperature limits. These results are shown in Figure\,\ref{Fig:DSMRC}(b) and indicate that the overall magnetocaloric performance in the series does improve as Mn content is increased since the refrigerant capacity increases nearly 5-fold throughout the series.

Often, \MSexp{} is used as a proxy for understanding the magnitude of the magnetocaloric effect, however in this case it is clear that other factors -- such as magnetostructural coupling -- are at play. For example, the saturation magnetization (Figure\,\ref{Fig:MHs}) of the $x$\,=\,0.75 and $x$\,=\,1.00 samples near 7\,T is approximately 1.12\,$\mu_B$ and 1.25\,$\mu_B$, respectively. However, the difference in their peak $\Delta S_M$ values nearly doubles from 3.51\,J\,kg\,$^{-1}$K\,$^{-1}$ to 5.67\,J\,kg\,$^{-1}$K\,$^{-1}$, respectively, indicating that in this case \MSexp{} is not an appropriate metric by which to understand magnetocaloric performance as quantified by $\Delta S_M$.

\subsection{First-Principles Calculations}
The experimental measurements of $\Delta S_M$ are compared to first-principles computational proxies for the $\Delta S_M$ figure of merit for magnetocaloric materials. Bocarsly \textit{et al}. \cite{Bocarsly2017} and others have demonstrated the predictive power of computational magnetic deformation proxy, $\Sigma_M$ \cite{Cooley2020}. The ``magnetic deformation proxy'' provides a computationally inexpensive indicator of a large change in magnetocaloric isothermal change in entropy, $\Delta S_M$ above a threshold of $\Sigma_M > 1.5\%$. $\Sigma_M$ is a measure of the deformation strain between the magnetic and nonmagnetic structures of the material. The reason for the strong correlation between $\Delta S_M$ and $\Sigma_M$ can be explained, in a general sense, by the role that coupled magnetic and structural degrees of freedom play in promoting phase transitions with a large change in entropy, and even latent heat in the case of a first-order magnetostructural phase transition.

In this study, the magnetic deformation proxy $\Sigma_M$ was computed for a representative set of structures for each Mn composition $x$ in Mn$_x$Co$_{1-x}$Cr$_2$O$_4$. In addition to $\Sigma_M$, previous studies have identified that the total magnetization at $T$\,=\,0 K, \MScomp{}, calculated from DFT, also correlates well with $\Delta S_M$, although to a lesser degree than $\Sigma_M$ for the materials database analyzed in the original study that tested the correlation of $\Delta S_M$ with different first-principles indicators \cite{Bocarsly2017}. In this context, \MScomp{} is defined as the ``net magnetization'' within the unit cell and can be computed from the difference between spin-up and spin-down electron densities produced by DFT calculations. Bocarsly and others have shown that for a representative sample of ferrimagnetic and antiferromagnetic materials, the product of the deformation proxy with the saturation magnetization, \MScomp{} $\Sigma_M$, in many cases provides a greater indication of a large change in entropy, $\Delta S_M$, compared to either \MScomp{} or $\Sigma_M$ individually \cite{UCSBmag}. 

\begin{figure}
    \includegraphics[scale=0.45]{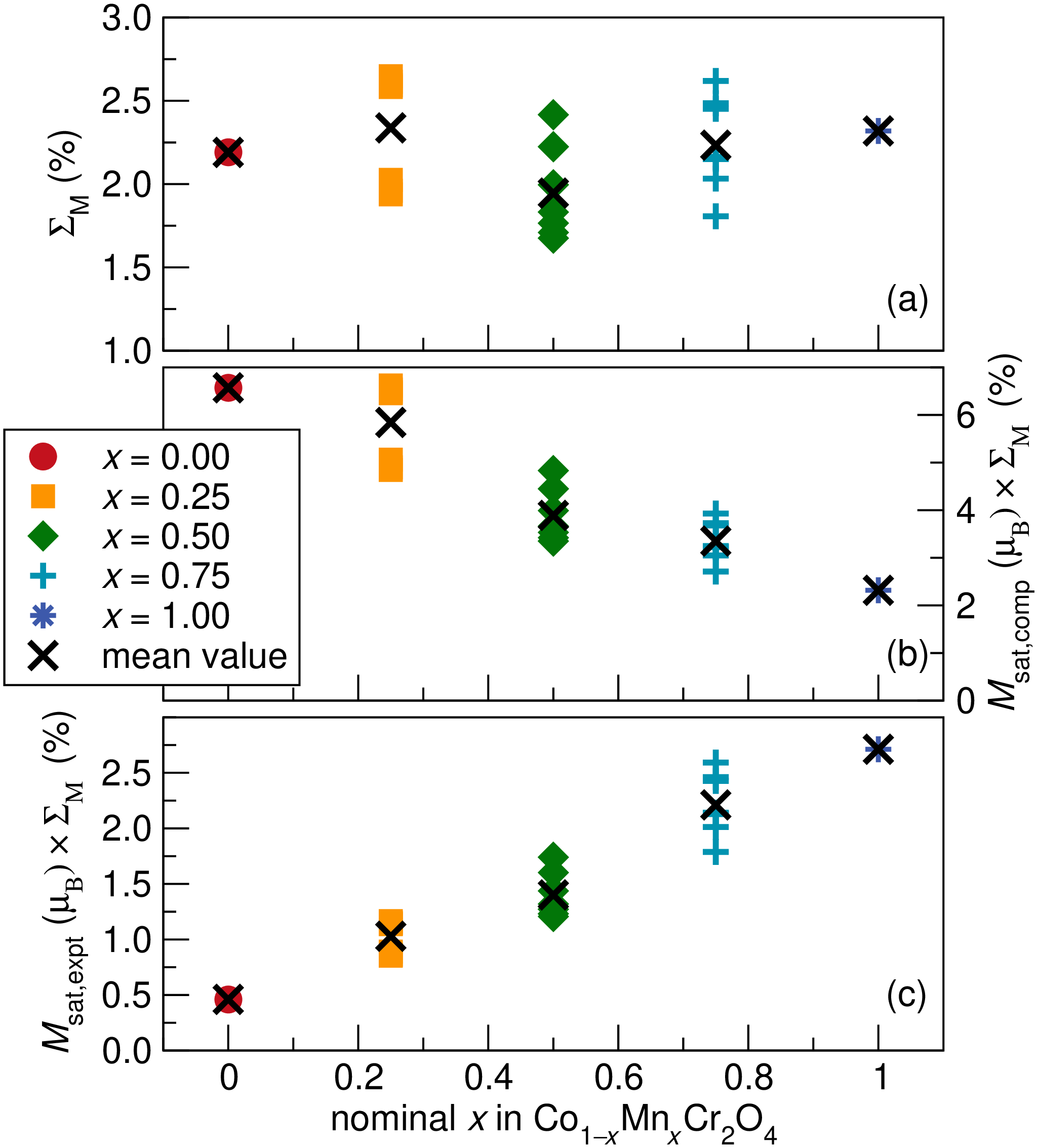}
    \caption{(a) The deformation proxy, $\Sigma_M$, versus Mn composition. (b) The computed net magnetization \MScomp{} ($\mu_B$ per f.u.) scaled by the deformation proxy, \MScomp{} $\Sigma_M$, versus Mn composition. (c) The experimentally measured saturation magnetization \MSexp{} ($\mu_B$ per f.u.) scaled by the deformation proxy, $M_{sat} \Sigma_M$, versus Mn composition. Colored data-points indicate the individual calculations for each enumerated structure. Each black X indicates the mean deformation value at each composition. Over $x = $ 0.25, 0.5, and 0.75, the maximum standard deviation in energy per atom values at each composition computed by VASP was 0.4 meV.}
    \label{fig:SigmaMComp}
\end{figure}

Figure \ref{fig:SigmaMComp}(a) includes the distribution of $\Sigma_M$ values at each composition. Due to the large spread of values associated with a different crystal structure, we are unable to make a conclusion regarding the trend of the deformation values versus composition. The reported standard deviation of energy values was found to be no more than 0.4 meV, normalized by the number of atoms in the unit cell. The maximum difference in average energy per atom output by VASP was 1 meV for each ferrimagnetic configuration for both CoCr$_2$O$_4$ and MnCr$_2$O$_4$. This small difference in energy values may be indicative of the frustrated nature of magnetism that has been confirmed experimentally in in CoCr$_2$O$_4$ and MnCr$_2$O$_4$ \cite{yoon_nmr_2010, PhysRevB.70.214434, Dey2014, Lin2017, Lawes2006, Yang2012} and theoretically using the classical theory of magnetic ground-states \cite{lyons_classical_1962, smart_netheory_1955}.

In addition to $\Sigma_M$, Figure \ref{fig:SigmaMComp}(b) shows the trend of the product of the computed net magnetization and the deformation proxy \MScomp{} $\Sigma_M$ versus composition. \MScomp{} is the net magnetization of the simulation cell output by VASP. These \MScomp{} values are in units of Bohr magnetons ($\mu_B$), normalized by the formula unit (f.u.). In this case, there is a clear downward trend, which would be expected for the collinear configuration that we chose to compute. \MScomp{} $\Sigma_M$ decreases with $x$, which is opposite to the trend that was observed from the experimental measurements. 

Compared to the DFT derived \MScomp{} $\Sigma_M$, Figure \ref{fig:SigmaMComp}(c) provides a plot of the experimentally measured \MSexp{} scaled by $\Sigma_M$ at each composition. These \MSexp{} values are also reported in Bohr magnetons, normalized by the formula unit. The upward trend with composition is due to the dominant behavior of the experimentally measured \MSexp{} values, which increase with Mn composition.

From the results of the computed $\Sigma_M$ and \MScomp{} $\Sigma_M$ versus Mn composition, it is clear that there is a large disagreement with the experimentally observed behavior. This is possibly due to the fact that $\Sigma_M$ simply quantifies the degree of magnetostructural coupling in a material, without treating the lattice and spin contributions to $\Delta S$ of the phase transition explicitly. The opposite trend of \MScomp{} with the experimental saturation magnetization, \MSexp{}, is stark, but can be described by the inconsistencies between \MSexp{} and \MScomp{}. The two are strictly comparable only at zero temperature. Even then, \MScomp{} is calculated under zero applied field, where \MSexp{} is not, by definition. The saturation magnetization and effective moment \MUeff{} are inherently temperature-dependent quantities. For example, the \MUeff{} is derived from the high temperature paramagnetic decay of the susceptibility according to the Curie-Weiss law. For this reason, Monte Carlo methods or a mean-field description is necessary in order to connect the DFT derived parameters to the experimentally measured thermodynamic quantities such as \MSexp{} and \MUeff{}. In addition, it has been shown that \MScomp{} doesn't correlate well with $\Delta S_M$ for a nonzero applied magnetic field \cite{Bocarsly2017}. 

In future studies of this spinel system, thermodynamic quantities versus temperature will be computed using a spin-lattice coupling Hamiltonian with parameters derived from density functional theory calculations. This model will allow for the quantification of the change in entropy due to structural, magnetic, and their coupling using Monte Carlo methods that directly quantify spin and phonon contributions to $\Delta S_M$.

\section{Conclusion}

We have studied the solid solution Co$_{1-x}$Mn$_x$Cr$_2$O$_4$ as a candidate for magnetocaloric applications by synthesizing using standard solid-state synthesis. Synchrotron X-ray diffraction measurements revealed a complete solid solution of  spinel samples. Magnetic measurements between 2\,K and 390\,K show the spiral spin-state at low temperatures transitioning to collinear ferrimagnetism at moderate temperatures then to paramagnetism at high temperatures, and the transition temperatures for each phase decrease monotonically as Mn content is increased. The maximum magnetic entropy change is also found to increase monotonically, from $-$1.67\,J\,kg\,$^{-1}$K\,$^{-1}$ for the Co end member to $-$5.63\,J\,kg\,$^{-1}$K\,$^{-1}$ for the Mn end member (for a field change of 0\,T to 5\,T). Overall, the tunability of this series and robust peak $\Delta S_M$ values in the range of 40\,K to 75\,K make this series attractive for cascaded hydrogen liquefaction systems relying on active magnetic refrigeration. 

The effect of variation in Mn/Co composition on maximum magnetic entropy difference, $\Delta S_M$, cannot be explained by DFT-computed magnetic deformation proxy values, $\Sigma_M$ \cite{Bocarsly2017}, but are more closely related to the trends in \MSexp{} across composition. This is \textit{not} the case for \MScomp{} from DFT, therefore, we suggest that there are crucial thermodynamic mechanisms that underlie the wide range of $\Delta S_M$ values across the series. We argue that this finite-temperature behavior cannot be resolved at the level of DFT alone \footnote{DFT is a fundamentally zero Kelvin theory.}, but requires finite temperature modeling approaches, such as Monte Carlo, or even Weiss mean-field models. 

\section{Acknowledgements} 
This work and facilities employed here were supported by  the National Science Foundation through the MRSEC Program NSF DMR 1720256 (IRG-1). 
The UCSB MRSEC is a member of the Materials Research Facilities Network (www.mrfn.org). Use of the Advanced Photon Source at Argonne National 
Laboratory was supported by the U. S. Department of Energy, Office of Science, Office of Basic Energy Sciences, under Contract No. DE-AC02-06CH11357. 
G.M. acknowledges support from the Department of Energy Computational Science Graduate Fellowship (DOE CSGF) under grant DE-SC0020347. 
M.K.H acknowledges support by the U.S. Department of Energy, Office of Science, Office of Basic Energy Sciences, Materials Sciences and Engineering 
Division under Contract No. DE-AC02-05-CH11231 (Materials Project program KC23MP).

\section{Appendix}

\subsection{Curie-Weiss Law for a Ferrimagnet}
\label{section:cw_sublattice_appendix}

\noindent The Weiss molecular field theory for magnetism \cite{weiss_hypothese_1907} is based on a mean-field approximation, and therefore neglects fluctuations that influence behavior below and near to the critical temperature. However, Curie-Weiss theory is useful for studying the behavior of magnets at temperatures above their paramagnetic transition temperature.
This theory was originally formulated for the ferromagnetic to paramagnetic order-disorder phase transition, however, it is possible to generalize this theory to the study of ferrimagnetism and antiferromagnetism, by treating the magnetic configuration as an antiparallel coupled arrangement of parallel (ferromagnetic) lattices. This first section follows the reasoning presented in the text of Kittel \cite{kittel_introduction_2005}.

For the sake of clarity, we will start from the system of equations for the Weiss molecular field coupling between the magnetization of A and B sublattices in a ferrimagnet, assuming a negative - AFM exchange between A and B sites,
\begin{align}
    {M}_A &= \frac{C_A}{\tempapp{}} \left( {B}_0 - \lambda_{AA} {M}_A - \lambda_{AB} {M}_B \right) \nonumber \\
    {M}_B &= \frac{C_B}{\tempapp{}} \left( {B}_0 - \lambda_{AB} {M}_A - \lambda_{BB} {M}_B \right)
    \label{eq:weiss_sys}
\end{align}
where ${M}_A$ \& ${M}_B$ and $C_A$ \& $C_B$ are the net magnetization and Curie-Weiss constants of the A and B sublattices, respectively. ${B}_0$ represents the applied field, and $\lambda_{AA}$, $\lambda_{BB}$, \& $\lambda_{AB}$ are the Weiss field constants. These Weiss field constants ($\lambda$'s) capture the net exchange interaction within and between the two sublattices. The latter of which is symmetric under both directions of the exchange (A $\rightarrow$ B and vice versa). Equations \ref{eq:weiss_sys} can be stated in matrix-vector form as Equation \ref{eq:weiss_matvec}.
\begin{widetext}
\begin{align}
  \left( T \cdot \bm I + \bm W \right)
  \begin{bmatrix}
    M_A \\ M_B
  \end{bmatrix}
  &=
  \left( T
  \begin{bmatrix}
    1 & 0 \\ 0 & 1 \\
  \end{bmatrix} \right.
  \left. +
  \begin{bmatrix}
    \lambda_{AA} C_A & \lambda_{AB} C_A \\ \lambda_{AB} C_B & \lambda_{BB} C_B \\
  \end{bmatrix}
  \right)
  \begin{bmatrix}
    M_A \\ M_B
  \end{bmatrix}
  = B_0
  \begin{bmatrix}
    C_A \\ C_B
  \end{bmatrix}
  \label{eq:weiss_matvec}
\end{align}
\end{widetext}
By inspecting Equation \ref{eq:weiss_matvec}, we see that for $B_0 = 0$, a nonzero solution for $M_A$ \& $M_B$ exists only if \cite{kittel_introduction_2005}
\begin{align}
  \det \left\{
  T \cdot \bm I + \bm W
  \right\} &= 0
  \label{eq:det_cond}
\end{align}
Therefore, the critical temperature of the system will correspond to the opposite sign of an eigenvalue of $\bm W$, specifically $T_C = \max\left\{ \eta_i \left[ -\bm W \right] \right\}$, where $\eta_i \left[ \bm A \right]$ are the eigenvalues of a matrix $\bm A$. From Equation \ref{eq:weiss_matvec}, we arrive at a generalized Curie-Weiss law, for $T > T_C$
\begin{align}
  \chi &= \frac{\partial \left(\sum_{i=1}^n M_i \right)}{\partial {B}_0 } = 
  \mathbf{1}^T \left[ T \cdot \bm I + \bm W \right]^{-1} \bm C
  \label{eq:cw_general}
\end{align}
For the two lattice case ($n=2$), \mbox{$\sum_{i=1}^n M_i = {M}_A + {M}_B$}, \mbox{$\bm 1^T = \left[1\ 1\right]$}, and \mbox{$\bm C^T = \left[C_A\ C_B\right]$}.

\subsubsection{Uncertainty quantification}

In order to fit Equation \ref{eq:cw_general} to experimental data, we used \CurveFitName{} \cite{virtanen2020scipy}. This optimization routine utilizes the Jacobian of the objective function in the minimization procedure, as well as for calculating the propagation of uncertainty. If one does not specify the \texttt{jacobian} argument explicitly, then \CurveFitName{} computes derivatives using finite differences (FD). If we use the default FD scheme, the uncertainty values of the parameters are estimated to be at least $10^4$. However, we find improvement if we supply the analytical derivative presented below in Equation \ref{eq:cw_general_deriv},
\begin{align}
  \frac{\partial}{\partial \varphi} \chi
  &= - \mathbf{1}^T \bm W_T^{-1} \frac{\partial \bm W}{\partial \varphi} \bm W_T^{-1} \bm C
  + \mathbf{1}^T \bm W_T^{-1} \frac{\partial \bm C}{\partial \varphi}, \nonumber \\
  \text{where} \quad 
  \bm W_T &= \left[ T \cdot \bm I + \bm W \right].
  \label{eq:cw_general_deriv}
\end{align}
In this expression, $\varphi$ represents a parameter of the model function, \mbox{$\varphi \in \left\{ C_A, C_B, \lambda_{AA}, \lambda_{BB}, \lambda_{AB} \right\}$}. If we provide the analytical Jacobian, all uncertainties fell below 1. This improvement is likely due to numerical artifacts that arise from using the FD scheme near to the singularity at the critical temperature.

\subsubsection{Simplified exchange}

We can simplify the Equation \ref{eq:weiss_matvec} by setting $\lambda_{AA} = \lambda_{BB} = 0$, and letting $\lambda = \lambda_{AB}$. This simplification yields the following result for susceptibility above the critical temperature $\tempapp{}_C$ \cite{kittel_introduction_2005}:
\begin{align}
    \label{eqn:sx_AB}
    \chi = \frac{\partial \left({M}_A + {M}_B \right)}{\partial {B}_0 } = \frac{\left( C_A + C_B \right)\tempapp{} - 2 \lambda C_A C_B}{\tempapp{}^2 - \tempapp{}_C^2}
\end{align}
Another consequence of this solution is that $\tempapp{}_C = \lambda \sqrt{C_A C_B}$ \cite{kittel_introduction_2005}.

\subsubsection{High temperature limit}

\begin{figure}[H]
    \includegraphics[width=0.9\linewidth]{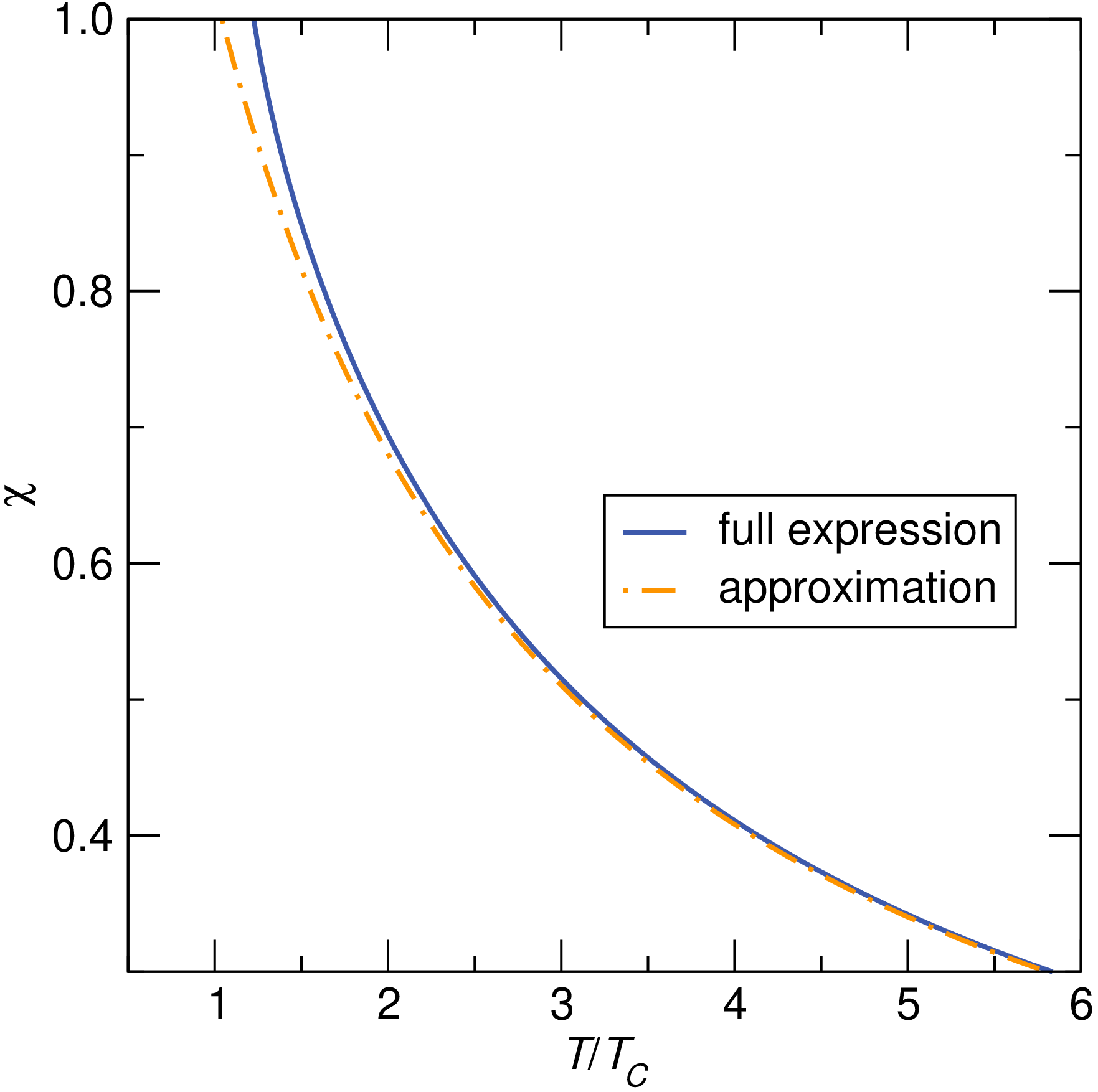}
    \caption{Comparison between the Curie-Weiss (CW) expressions for $\chi$, where the full expression (Equation \ref{eqn:sx_AB}) is plotted against the approximate high-temperature CW law (Equation \ref{eqn:CW_FiM_simple}). The parameters, $\lambda$, $C_A$, and $C_B$ are arbitrary dimensionless constants that are chosen to justify this approximation.}
    \label{fig:chi_approx}
\end{figure}

\noindent Continuing from Equation \ref{eqn:sx_AB}, we can obtain the following factorization
\begin{align}
    \chi &= \frac{\partial \left({M}_A + {M}_B \right)}{\partial {B}_0 } \nonumber \\
    &= \frac{C_A + C_B}{\tempapp{} + \tempapp{}_C} \cdot \frac{\tempapp{} - 2 \lambda \frac{C_A C_B}{C_A + C_B}}{\tempapp{}-\tempapp{}_C} \\
    &= \frac{C_A + C_B}{\tempapp{} + \tempapp{}_C} \cdot \frac{\tempapp{} - \rho \cdot \tempapp{}_C}{\tempapp{}-\tempapp{}_C}
\end{align}
where $\rho$ is the ratio between the geometric and arithmetic mean between $C_A$ and $C_B$,
\begin{align}
\rho= \frac{C_A^{1/2} C_B^{1/2}}{\frac{1}{2}\left(C_A + C_B\right)}.
\end{align}
Therefore, $\rho = 1$ if $C_A = C_B$. If we examine the high-temperature limit in which $\tempapp{} >> \tempapp{}_C$, we can make the following simplification
\begin{align}
    \label{eqn:highT_assumption}
    \lim_{\tempapp{} \rightarrow \infty} \frac{\tempapp{} - \rho \cdot \tempapp{}_C}{\tempapp{}-\tempapp{}_C} = 1
\end{align}
Under approximation, we arrive at a more compact form of the CW law for a ferrimagnet
\begin{align}
    \label{eqn:CW_FiM_simple}
    \chi \approx \frac{C_A + C_B}{\tempapp{} + \tempapp{}_C},
\end{align}
which is applicable for temperatures significantly larger than the critical temperature. An illustration of this is shown in Figure \ref{fig:chi_approx}.

\subsubsection{Relation to effective moment}

\noindent The effective moment can be expressed in terms of the Curie-Weiss law using Weiss mean-field theory \cite{kittel_introduction_2005}
\begin{align}
    \label{eqn:CW_effmom}
    \chi &= \frac{Np^2\mu_b^2}{3 k_B (\tempapp{}-\tempapp{}_C)} = \frac{N \mueff{}^2 }{3 k_B (\tempapp{}-\tempapp{}_C)} = \frac{C}{\tempapp{}-\tempapp{}_C} \nonumber \\
    \mueff{}^2 &= \frac{3k_B}{N} C
\end{align}
$N$ is the number of atoms per unit volume \cite{kittel_introduction_2005}.

Combining Equations \ref{eqn:CW_FiM_simple} and \ref{eqn:CW_effmom}, we arrive at the following relationship between the ``net" effective moment and the effective moment for each sublattice (A \& B):
\begin{align}
    \label{eqn:net_effmom}
    C &\approx C_A + C_B = \frac{N_a\nonumber \mu_a^2}{3k_B} + \frac{N_b \mu_b^2}{3k_B} \nonumber \\
    \mueff{}^2 &\approx \frac{N_a}{N} \mu_a^2 + \frac{N_b}{N} \mu_b^2
\end{align}
This relationship allows us to approximate, at $T>T_C$, the effective magnetic moment of a sample from a weighted sum of the \MUeff{} of the constituent magnetic sublattices.

\subsubsection{Spinel system \GenSpinel{}}

\noindent Generalizing Equation \ref{eqn:net_effmom} to the spinel system \GenSpinel{}, we can use the respective chemical compositions of each element to ascertain their respective composition fractions $N_i/N$, where $N_i$ denotes the number of atoms of species $i$ and $N$ is the sum of the total number of atoms per unit volume:
\begin{align}
    \label{eqn:net_effmom_spinel}
    \mueff{}^2 &\approx \frac{N_a}{N} \mu_a^2 + \frac{N_c}{N} \mu_c^2 + \frac{N_b}{N} \mu_b^2 \nonumber \\
    \mueff{}^2 &\approx x \mu_a^2 + (1-x) \mu_c^2 + 2 \mu_b^2
\end{align}
Once again, this is using a high temperature approximation in which $\tempapp{}>>\tempapp{}_C$.

\bibliography{MnCoSpinel}

\end{document}